\documentstyle[12pt,epsfig]{article}

\newcommand{\agt}{\,\rlap{\lower 3.5 pt \hbox{$\mathchar \sim$}} \raise 1pt
 \hbox {$>$}\,}
\newcommand{\alt}{\,\rlap{\lower 3.5 pt \hbox{$\mathchar \sim$}} \raise 1pt
 \hbox {$<$}\,}

\catcode`@=11
\newcount\@tempcntc
\def\@citex[#1]#2{\if@filesw\immediate\write\@auxout{\string\citation{#2}}\fi
  \@tempcnta\z@\@tempcntb\m@ne\def\@citea{}\@cite{\@for\@citeb:=#2\do
    {\@ifundefined
       {b@\@citeb}{\@citeo\@tempcntb\m@ne\@citea\def\@citea{,}{\bf ?}\@warning
       {Citation `\@citeb' on page \thepage \space undefined}}%
    {\setbox\z@\hbox{\global\@tempcntc0\csname b@\@citeb\endcsname\relax}%
     \ifnum\@tempcntc=\z@ \@citeo\@tempcntb\m@ne
       \@citea\def\@citea{,}\hbox{\csname b@\@citeb\endcsname}%
     \else
      \advance\@tempcntb\@ne
      \ifnum\@tempcntb=\@tempcntc
      \else\advance\@tempcntb\m@ne\@citeo
      \@tempcnta\@tempcntc\@tempcntb\@tempcntc\fi\fi}}\@citeo}{#1}}
\def\@citeo{\ifnum\@tempcnta>\@tempcntb\else\@citea\def\@citea{,}%
  \ifnum\@tempcnta=\@tempcntb\the\@tempcnta\else
   {\advance\@tempcnta\@ne\ifnum\@tempcnta=\@tempcntb \else \def\@citea{--}\fi
    \advance\@tempcnta\m@ne\the\@tempcnta\@citea\the\@tempcntb}\fi\fi}
\catcode`@=12
\begin{document}

\title{\vskip-3cm{\baselineskip14pt
\centerline{\normalsize DESY 99-189\hfill ISSN 0418-9833}
\centerline{\normalsize MPI/PhT/99-59\hfill}
\centerline{\normalsize hep-ph/0002058\hfill}
\centerline{\normalsize December 1999\hfill}
}
\vskip1.5cm
Virtual Sfermion Effects on Vector-Boson Pair Production at $e^+e^-$ Colliders
}
\author{
{\sc A.A. Barrientos Bendez\'u,$^1$ K.-P.O. Diener,$^2$ B.A. Kniehl$^1$}\\
{\normalsize $^1$ II. Institut f\"ur Theoretische Physik, Universit\"at
Hamburg,}\\
{\normalsize Luruper Chaussee 149, 22761 Hamburg, Germany}\\
{\normalsize $^2$ Max-Planck-Institut f\"ur Physik
(Werner-Heisenberg-Institut),}\\
{\normalsize F\"ohringer Ring 6, 80805 Munich, Germany}}

\date{}

\maketitle

\thispagestyle{empty}

\begin{abstract}
We study the quantum effects on vector-boson pair production in $e^+e^-$ 
annihilation induced by the sleptons and squarks of the minimal 
supersymmetric extension of the standard model (MSSM) in the one-loop
approximation.
We list full analytic results, and quantitatively analyze the resulting
deviation from the standard-model prediction of $e^+e^-\to W^+W^-$ for the
supergravity-inspired MSSM.
The latter can be rendered small throughout the whole parameter space by an
appropriate choice of renormalization scheme.

\medskip

\noindent
PACS numbers: 12.60.Jv, 13.10.+q, 14.80.Ly
\end{abstract}

\newpage

The production of $W$-boson pairs in $e^+e^-$ annihilation offers a unique
opportunity to probe the nonabelian gauge structure of the standard model (SM)
at the tree level, which manifests itself in a distinctive cancellation
between the $s$-, $t$-, and $u$-channel scattering amplitudes.
This process is being studied experimentally with high precision at the CERN
Large Electron-Positron Collider (LEP2) in the centre-of-mass (CM) energy
range $2M_W\alt\sqrt s\alt205$~GeV.
At the same time, the related reactions $e^+e^-\to\gamma\gamma,\gamma Z,ZZ$
are being measured there, too.
With a future $e^+e^-$ linear supercollider, such as JLC, NLC, or TESLA, these
measurements can be extended to higher energies, way up to the TeV range, and
rendered more precise.
On the theoretical side, enormous effort has been invested into the
computation of the one-loop radiative corrections to the cross sections of
these processes in the SM, both for on- \cite{boh,sac} and off-shell
\cite{den} vector bosons, and useful low- \cite{dit} and high-energy
\cite{bee} approximations have been elaborated; for a comprehensive review,
see Ref.~\cite{wbe}.

Significant deviations of the measured cross sections from their SM
predictions could signal physics beyond the SM.
Since the gauge couplings of the electron are so tightly constrained by
low-energy and LEP1 data, such deviations should mainly originate from the
triple gauge-boson couplings (TGC's).
Generic parameterizations of the so-called anomalous TGC's were introduced and
applied to the processes $e^+e^-\to W^+W^-,ZZ$ in Ref.~\cite{gae}.
In order to explain the physical origin of anomalous TGC's, it is necessary to
consider specific new-physics scenarios.
From the theoretical point of view, renormalizable extensions of the SM are
most satisfactory.
As a rule, the deviations are then induced through loop effects of new
particles, which affect not only the TGC's, but also the vector-boson 
propagators and the renormalizations of the parameters and wave functions of 
the tree-level amplitudes.
Such deviations were investigated at the one-loop level in
Refs.~\cite{kat,die} for models with a modified lepton sector including
Majorana neutrinos and were found to be generally small.
In Ref.~\cite{arh}, the one-loop radiative corrections to the anomalous 
parameters $\Delta\kappa_V$ and $\lambda_V$ of the $VW^+W^-$ TGC's, with
$V=\gamma,Z$, were studied in the minimal supersymmetric extension of the SM
(MSSM), using the pinch technique to render them gauge independent.
The sfermion contributions were found to generally dominate the Higgs and 
gaugino contributions.
This may be understood by observing that mass splittings between the up and
down components of the sfermion doublets give rise to significant
contributions \cite{arh} and that the sfermions come in large numbers, due to
their multiplicities in flavour and colour.

In this letter, we calculate the sfermion contributions to the cross sections
of $e^+e^-\to V_1V_2$, with $V_1V_2=\gamma\gamma,\gamma Z,ZZ,W^+W^-$, at one
loop in the MSSM.
Preliminary results of this study were published in Ref.~\cite{hem}.
Our calculation proceeds along the lines of Ref.~\cite{die}, which gives
full analytic results.
We use the conventions of Ref.~\cite{die} and list only those formulas which
need to be substituted therein.
In a way, our analysis extends Ref.~\cite{rhe}, where the sfermion-induced
radiative corrections to the processes $e^+e^-\to Zh^0$ and $Z\to\gamma h^0$,
with $h^0$ being the lightest CP-even Higgs boson, were calculated at one loop
in the MSSM.
The authors of Ref.~\cite{arh} did not list analytic results that could be
compared with ours.
In Refs.~\cite{ala1,ala2,ala3}, the one-loop radiative correction to
$e^+e^-\to W^+W^-$ in the MSSM was considered for the full supersymmetric
particle spectrum, under the simplistic assumption that the mass matrix of
each sfermion flavour is proportional to the unit matrix, so that the two weak 
eigenstates are mass eigenstates with a common mass.
We shall compare the sfermion loop correction of Refs.~\cite{ala1,ala2,ala3}
with our result below.
As we shall see later, the size of the correction is significantly affected
by the sfermion mass splittings.

The Higgs sector of the MSSM is made up by two complex Higgs isodoublet of
opposite hypercharge and accommodates five physical Higgs bosons:
the neutral CP-even $h^0$ and $H^0$ bosons, the neutral CP-odd $A^0$ boson,
and the charged $H^\pm$-boson pair.
At the tree level, it has two free parameters, which are usually taken to be
the mass $m_A$ of the $A^0$ boson and the ratio $\tan\beta=v_2/v_1$ of the
vacuum expectation values of the two Higgs doublets.
For each of these Higgs bosons and each SM fermion and gauge boson there is a 
supersymmetric partner.
Thus, the spectrum of states is more than doubled if one passes from the SM to
the MSSM, which gives rise to a proliferation of parameters and weakens the
predictive power of the theory.
A canonical method to reduce the number of parameters is to embed the MSSM
into a grand unified theory (GUT), e.g., a suitable supergravity (SUGRA)
model, in such a way that it is recovered in the low-energy limit.
The MSSM thus constrained is described by the following parameters at the GUT 
scale, which come in addition to $\tan\beta$ and $m_A$: the universal scalar
mass $m_0$, the universal gaugino mass $m_{1/2}$, the trilinear Higgs-sfermion
coupling $A$, the bilinear Higgs coupling $B$, and the Higgs-higgsino mass
parameter $\mu$.
Notice that $m_A$ is then not an independent parameter anymore, but it is 
fixed through the renormalization group equation.
The number of parameters can be further reduced by making additional
assumptions.
Unification of the tau and bottom Yukawa couplings at the GUT scale leads to a
correlation between $m_t$ and $\tan\beta$.
Furthermore, if the electroweak symmetry is broken radiatively, then $B$ and
$\mu$ are determined up to the sign of $\mu$.
Finally, it turns out that the MSSM parameters are nearly independent of the
value of $A$, as long as $|A|\alt500$~GeV at the GUT scale.

We now present our analytic results.
We denote the four-momenta of $e^+$, $e^-$, and the two produced vector
bosons, $V_1$ and $V_2$, by $p_+$, $p_-$, $k_1$, and $k_2$, and define the
Mandelstam variables as $s=(p_++p_-)^2$, $t=(p_+-k_1)^2$, and $u=(p_+-k_2)^2$.
Neglecting the electron mass, we have $s+t+u=M_1^2+M_2^2$, where $M_1$ and
$M_2$ are the masses of $V_1$ and $V_2$, respectively.
In this limit, also the $s$-channel contributions due to Higgs-boson exchanges
vanish.
Because each of the four processes $e^+e^-\to V_1V_2$ has more than one
tree-level diagram, it is convenient to introduce helicity amplitudes
${\cal M}^\kappa(\lambda_1,\lambda_2,s,t)$, where $\kappa$, $\lambda_1$, and
$\lambda_2$ denote the helicities of $e^-$, $V_1$, and $V_2$ in the CM frame,
respectively.
The $e^+$ helicity is then $-\kappa$.
The helicity amplitudes ${\cal M}^\kappa$ can be decomposed into the standard
matrix elements ${\cal M}_i^\kappa$ ($i=0,\ldots,9$) \cite{sac}, which are
written down in Appendix~B of Ref.~\cite{die}.
In addition to those for $i=0,\ldots,3$, which already appear at the tree
level, we only need ${\cal M}_9^\kappa$ for the present analysis.

The tree-level cross sections of $e^+e^-\to V_1V_2$ are well known and may be 
found in Eq.~(3.1) of Ref.~\cite{die}.
The sfermion-induced one-loop corrections receive contributions from diagrams
containing self-energy corrections, vertex corrections, and counterterm
insertions.
We work in the Fermi-constant ($G_F$) formulation of the electroweak on-shell
renormalization scheme, which is explained in the context of Eq.~(4.1) in
Ref.~\cite{die}.
Specifically, starting from the results in the pure on-shell renormalization
scheme, which uses Sommerfeld's fine-structure constant $\alpha$ and the
physical particle masses as basic parameters, we fix
$\alpha=\sqrt2G_F\sin^2\theta_wM_W^2/\pi$, where $\theta_w$ is the weak
mixing angle, and supplement the radiative corrections with the term
$-2\Delta r$, where $\Delta r$ \cite{sir} contains those radiative corrections
to the muon lifetime which the SM or its extensions introduce on top of the
purely photonic corrections from within the Fermi model.
The sfermion contribution to $\Delta r$ in the MSSM was examined in
Ref.~\cite{rhe}.
All the formulas listed in Section~III and Appendix~D of Ref.~\cite{die} carry
over to the sfermion case, except for Eqs.~(3.5) and (3.8), which give the
transverse parts of the vector-boson vacuum polarizations $\Pi_T^{V_1V_1}$ and
the proper vertex corrections $\delta{\cal M}_V^\kappa$, respectively.
The relevant Feynman rules for the MSSM sfermion sector are summarized in 
Appendix~A of Ref.~\cite{rhe}.
For each fermion flavour $Q=U,D$, where $U=\nu_e,\nu_\mu,\nu_\tau,u,c,t$ and
$D=e,\mu,\tau,d,s,b$, there is a corresponding sfermion flavour, denoted by a
tilde.
Except for the sneutrinos, which we assume to be left handed, $\tilde Q$ comes
in two mass eigenstates $a=1,2$.
The masses $M_{\tilde Qa}$ of the sfermions and their trilinear and quartic
couplings, $\tilde V_{Q_aQ_b^\prime}^{V_i}$ and
$\tilde U_{Q_aQ_b^\prime}^{V_iV_j}$, respectively, to the vector bosons
$V_i=\gamma,Z,W$ are defined in Appendix~A of Ref.~\cite{rhe}.
In the absence of flavour-changing neutral currents, we have $Q=Q^\prime$ in
$\tilde V_{Q_aQ_b^\prime}^{V_i}$ and $\tilde U_{Q_aQ_b^\prime}^{V_iV_j}$ if
$V_i,V_j=\gamma,Z$, which explains the notation $\tilde V_{Qab}^{V_i}$ and
$\tilde U_{Qab}^{V_iV_j}$ used in Ref.~\cite{rhe}.
As in Ref.~\cite{rhe}, we neglect the Cabibbo-Kobayashi-Maskawa mixing, so
that we may write $\tilde V_{UaDb}^W$ and $\tilde U_{Qab}^{WW}$.
The sfermion contributions to the $\Pi_T^{V_1V_2}$ functions read \cite{rhe}
\begin{eqnarray}
\Pi_T^{V_1V_2}(p^2)&=&\frac{1}{48\pi^2}\sum_{Q,a,b}N_{\mathrm{col}}^Q
\tilde V_{Q_aQ_b}^{V_1}\tilde V_{Q_bQ_a}^{V_2}
\left\{\left[s-2\left(M_{\tilde Q_a}^2+M_{\tilde Q_b}^2\right)
+\frac{\left(M_{\tilde Q_a}^2-M_{\tilde Q_b}^2\right)^2}{s}\right]
\right.\nonumber\\
&&{}\times B_0\left(p,M_{\tilde Q_a},M_{\tilde Q_b}\right)
+M_{\tilde Q_a}^2\left(2-\frac{M_{\tilde Q_a}^2-M_{\tilde Q_b}^2}{s}\right)
B_0\left(0,M_{\tilde Q_a},M_{\tilde Q_a}\right)
\nonumber\\
&&{}+\left.
M_{\tilde Q_b}^2\left(2-\frac{M_{\tilde Q_b}^2-M_{\tilde Q_a}^2}{s}\right)
B_0\left(0,M_{\tilde Q_b},M_{\tilde Q_b}\right)
+\frac{2}{3}s-\frac{\left(M_{\tilde Q_a}^2-M_{\tilde Q_b}^2\right)^2}{s}
\right\},\quad
\label{eq:pi}
\end{eqnarray}
where $p$ is the external four-momentum, $N_{\mathrm{col}}^Q=1$ (3) for
sleptons (squarks), and the standard two-point scalar function $B_0$ is
defined in Eq.~(C.2) of Ref.~\cite{die}.
If $V_1=W^-$ and $V_2=W^+$, then $Q_a=U_a$, $Q_b=D_b$, and it is summed over
$(U,D)$ instead of $Q$.
The sfermion contribution to $\delta{\cal M}_V^\kappa$ is found to be
\begin{eqnarray}
\delta{\cal M}_V^\kappa&=&-\frac{1}{2\pi^2}\sum_{B=\gamma,Z}
\frac{g_{eeB}^\kappa}{s-M_B^2}\sum_{Q,a,b,c}N_{\mathrm{col}}^Q
\left(\tilde V_{Q_cQ_b}^B\tilde V_{Q_bQ_a}^{V_1}\tilde V_{Q_aQ_c}^{V_2}
-\tilde V_{Q_cQ_a}^{V_2}\tilde V_{Q_aQ_b}^{V_1}\tilde V_{Q_bQ_c}^B\right)
\nonumber\\
&&{}\times\left[{\cal M}_1^\kappa\left(C_2^0+C_3^{01}+C_3^{02}\right)
+{\cal M}_2^\kappa C_3^{01}+{\cal M}_3^\kappa C_3^{02}
-{\cal M}_9^\kappa\left(C_2^{12}+C_3^{12}+C_3^{21}\right)\right],
\label{eq:dm}
\end{eqnarray}
where the $C$ functions are the Lorentz coefficients of the standard
three-point tensor integrals defined in Eq.~(C4) of Ref.~\cite{die}.
In Eq.~(\ref{eq:dm}), we have suppressed their common argument
$\left(k_1,-k_2,M_{\tilde Q_a},M_{\tilde Q_b},M_{\tilde Q_c}\right)$.
Deviating from Ref.~\cite{die}, the electron gauge couplings appearing in
Eq.~(\ref{eq:dm}) are defined as
\begin{equation}
g_{ee\gamma}^\pm=e,\qquad
g_{eeZ}^+=-\frac{e\sin\theta_w}{\cos\theta_w},\qquad
g_{eeZ}^-=\frac{e}{\cos\theta_w\sin\theta_w}
\left(\frac{1}{2}-\sin^2\theta_w\right),
\end{equation}
where $e=\sqrt{4\pi\alpha}$.
We caution the reader that the Feynman rules used in Refs.~\cite{die,rhe}
differ in the sign of $\sin\theta_w$.
Consequently, we need to multiply the expression for $\Pi_T^{\gamma Z}$ in
Eq.~(\ref{eq:pi}) with an extra minus sign when we insert it into the relevant
formulas, Eqs.~(3.6) and (D2), of Ref.~\cite{die}.
Notice that Eq.~(\ref{eq:dm}) includes the contributions from both the direct
and crossed triangle diagrams, which are proportional to
$\tilde V_{Q_cQ_b}^B\tilde V_{Q_bQ_a}^{V_1}\tilde V_{Q_aQ_c}^{V_2}$ and
$\tilde V_{Q_cQ_a}^{V_2}\tilde V_{Q_aQ_b}^{V_1}\tilde V_{Q_bQ_c}^B$, 
respectively.
If $V_1=W^-$ and $V_2=W^+$, then the first term contributes for $Q_a=U_a$,
$Q_b=D_b$, and $Q_c=D_c$ and the second one for $Q_a=D_a$, $Q_b=U_b$, and
$Q_c=U_c$.

At this point, we should compare our results with those published in
Refs.~\cite{ala1,ala2,ala3}.
To that end, we put $M_{\tilde Q_1}=M_{\tilde Q_2}$ and nullify the mixing
angle relating the weak and mass eigenstates for each sfermion flavour
$\tilde Q$.
Then, our Eq.~(\ref{eq:pi}) agrees with Eqs.~(C7), (D5), (E9), and (F8) in 
Ref.~\cite{ala1}, up to an overall minus sign, if we eliminate the factor 1/2
multiplying $T_{3f}^i$ in Eq.~(D5) and the sum over $i$ in Eq.~(F8).
As for the $\gamma W^+W^-$ vertex correction, our Eq.~(\ref{eq:dm}) is in
accordance with Eqs.~(65) and (B13)--(B18) in Ref.~\cite{ala2} if we replace
the first two appearances of $C_{36}$ in Eq.~(B14) by $C_{35}$, substitute
$C_{24}$ in Eq.~(B18) by $C_{22}$, and include an overall minus sign in
Eqs.~(B17) and (B18).
As for the $ZW^+W^-$ vertex correction, we find agreement with Eqs.~(97) and
(C48)--(C53) in Ref.~\cite{ala2} if we alter the overall signs of Eqs.~(C52)
and (C53).

Now, we explore the phenomenological implications of our results.
We concentrate on the case of $e^+e^-\to W^+W^-$ because, for
$\sqrt s\agt180$~GeV, it has the largest cross section of the four processes
under consideration and it is the only one involving TGC's at the tree level 
in the SM.
The SM input parameters for our numerical analysis are taken to be
$G_F=1.16639\cdot10^{-5}$~GeV$^{-2}$ \cite{pdg}, $m_W=80.385$~GeV,
$m_Z=91.1871$~GeV, $m_t=174.3$~GeV \cite{eww}, and $m_b=4.7$~GeV.
We vary $\tan\beta$ and $m_A$ in the ranges $1<\tan\beta<35\approx m_t/m_b$
and 100~GeV${}<m_A<600$~TeV, respectively.
As for the GUT parameters, we choose $m_{1/2}=150$~GeV, $A=0$, and $\mu<0$,
and tune $m_0$ so as to be consistent with the desired value of $m_A$.
All other MSSM parameters are then determined according to the SUGRA-inspired
scenario as implemented in the program package SUSPECT \cite{djo}.
We checked that the results obtained from the program package ISAJET~7.49
\cite{bae}, where the electroweak-symmetry-breaking scale is fixed to be
$Q=\sqrt{M_{\tilde t_L}M_{\tilde t_R}}$, agree with those from SUSPECT within
typically 5\% or less if the same scale convention is implemented in the
latter.
In our analysis, we adopt the SUSPECT default value $Q=M_Z$.
We do not impose the unification of the tau and bottom Yukawa couplings at the
GUT scale, which would just constrain the allowed $\tan\beta$ range without
any visible effect on the results for these values of $\tan\beta$.
We exclude solutions which do not comply with the present experimental lower
mass bounds of the sfermions, charginos, neutralinos, and Higgs bosons 
\cite{ruh}.

\begin{figure}[ht]
\begin{center}
\centerline{\epsfig{figure=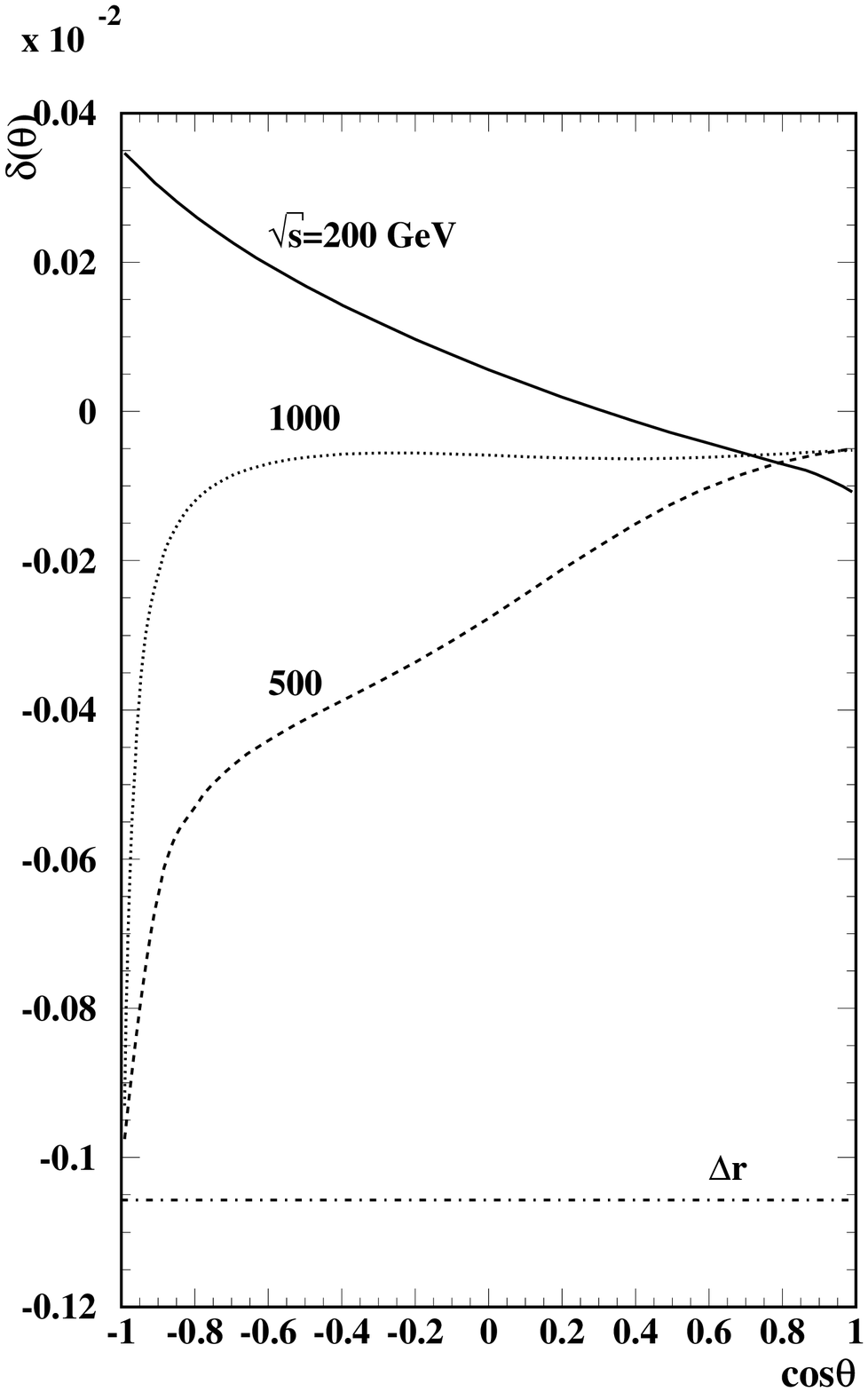,height=18cm}}
\caption{Relative correction $\delta(\theta)$ to the differential cross
section $d\sigma/d\cos\theta$ of $e^+e^-\to W^+W^-$, for
$\protect\sqrt s=200$, 500, and 1000~GeV, and contribution to $\Delta r$ due
to the sfermions in the SUGRA-inspired MSSM with $\tan\beta=10$ and
$m_A=250$~GeV, as functions of $\cos\theta$.}
\label{fig:one}
\end{center}
\end{figure}

\begin{figure}[ht]
\begin{center}
\centerline{\epsfig{figure=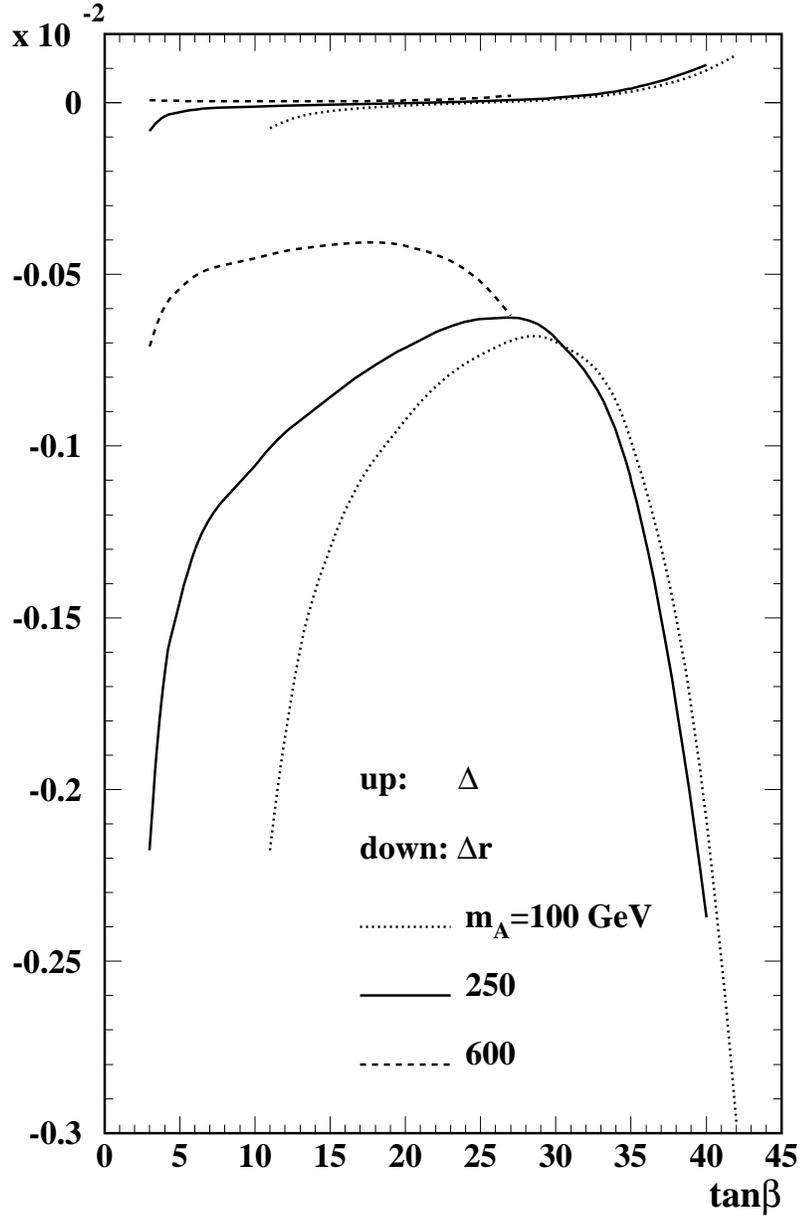,height=18cm}}
\caption{Relative correction $\Delta$ to the total cross section $\sigma$ of
$e^+e^-\to W^+W^-$, for $\protect\sqrt s=200$~GeV, and contribution to
$\Delta r$ due to the sfermions in the SUGRA-inspired MSSM with $m_A=100$,
250, and 600~GeV, as functions of $\tan\beta$.}
\label{fig:two}
\end{center}
\end{figure}

\begin{figure}[ht]
\begin{center}
\centerline{\epsfig{figure=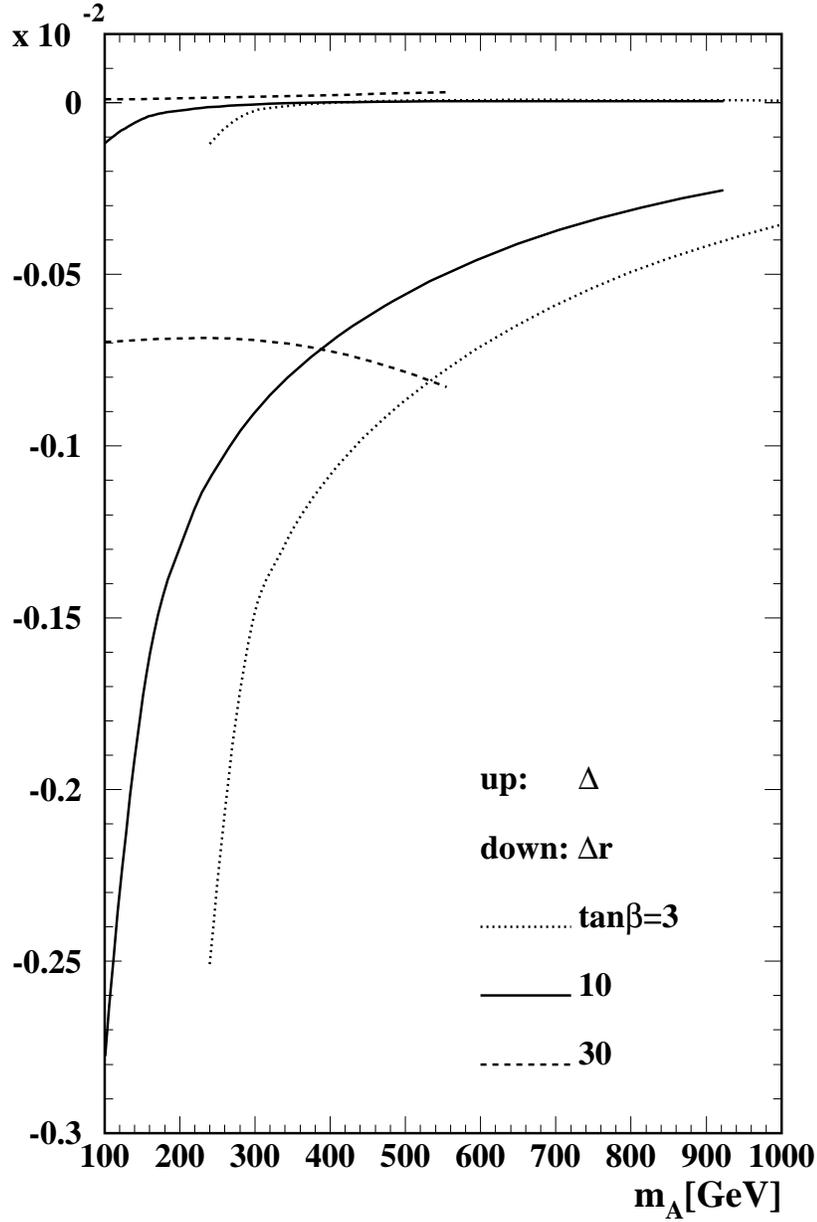,height=18cm}}
\caption{Relative correction $\Delta$ to the total cross section $\sigma$ of
$e^+e^-\to W^+W^-$, for $\protect\sqrt s=200$~GeV, and contribution to
$\Delta r$ due to the sfermions in the SUGRA-inspired MSSM with
$\tan\beta=3$, 10, and 30, as functions of $m_A$.}
\label{fig:three}
\end{center}
\end{figure}

In Fig.~\ref{fig:one}, the sfermion-induced correction $\delta(\theta)$ in the
relationship $d\sigma/d\cos\theta={}$\break
$(d\sigma/d\cos\theta)_{\mathrm{Born}}[1+\delta(\theta)]$ 
between the one-loop-corrected and tree-level cross sections of
$e^+e^-\to W^+W^-$ is shown as a function of the scattering angle $\theta$,
enclosed between the $e^+$ and $W^+$ three-momenta in the CM frame, for
$\sqrt s=200$, 500 and 1000~GeV assuming $\tan\beta=10$ and $m_A=250$~GeV.
We observe that $\delta(\theta)$ has a typical size of order 0.1\% or less and
can be of either sign.
In the backward direction, it strongly depends on the CM energy, while the
energy dependence is rather feeble in the forward direction.
We emphasize that the smallness of $\delta(\theta)$ is a special feature of
the $G_F$ scheme.
In the $\alpha$ scheme, the correction is given by $\delta(\theta)+2\Delta r$
and thus shifted to negative values because we have $\Delta r\approx-0.11\%$,
as indicated in Fig.~\ref{fig:one}.
Next, we study the correction $\Delta$ to the integrated cross section,
defined by $\sigma=\sigma_{\mathrm{Born}}(1+\Delta)$, for $\sqrt s=200$~GeV.
In Fig.~\ref{fig:two}, the $\tan\beta$ dependences of $\Delta$ and $\Delta r$
are shown for $m_A=100$, 250, and 600~GeV, while, in Fig.~\ref{fig:three},
the $m_A$ dependences are shown for $\tan\beta=3$, 10, and 30.
These dependences are implicit in the sense that our formulas for $\Delta$ and
$\Delta r$ do not contain $\tan\beta$ or $m_A$.
In fact, $\Delta$ and $\Delta r$ only depend on $\tan\beta$ or $m_A$ via the
sfermion masses and gauge couplings, the latter being affected through the
mixing angles which rotate the weak eigenstates of the sfermion into their
mass eigenstates.
We note that the SUGRA-inspired MSSM with our choice of input parameters does
not permit $\tan\beta$ and $m_A$ to be simultaneously small, due to the 
experimental selectron mass lower bound \cite{ruh}.
This explains why the curves for $m_A=100$~GeV in Fig.~\ref{fig:two} only
start at $\tan\beta\approx11$ and those for $\tan\beta=3$ in
Fig.~\ref{fig:three} at $m_A\approx240$~GeV.
For large $m_A$, the experimental $m_h$ lower bound \cite{ruh} enforces
$\tan\beta\agt3$.
On the other hand, the experimental lower bounds on the chargino and 
neutralino masses \cite{ruh} induce an upper limit on $\tan\beta$, which
depends on $m_A$.
From Fig.~\ref{fig:two} we observe that the $\tan\beta$ dependence of
$\Delta r$ for fixed $m_A$ is modest for intermediate values of $\tan\beta$,
while $\Delta r$ increases in magnitude towards the edges of the allowed
$\tan\beta$ range.
The stau and tau-sneutrino contributions dominate for large $\tan\beta$ and
small $m_A$, while the sbottom and stop contributions dominate for small
$\tan\beta$ and large $m_A$.
The contributions due to the sfermions of the first and second generations are
insignificant for all values of $\tan\beta$ and $m_A$.
It is interesting to investigate the mixings between the left- and
right-handed components of the charged sfermions in the third generation.
The mixing is strongest for stop, especially for small $\tan\beta$ and small
$m_A$.
For stau and sbottom, the mixings are generally feeble for large $\tan\beta$,
independently of $m_A$.
The magnitude of $\Delta r$ may reach several tenths of percent if $m_A$ is 
small to medium and $\tan\beta$ is close to its lower or upper limits.
For $\tan\beta\approx30$, $\Delta r$ is almost independent of $m_A$, while
for smaller (larger) values of $\tan\beta$, the size of $\Delta r$
monotonically decreases (increases) as $m_A$ increases.
These features are also nicely illustrated in Fig.~\ref{fig:three}.
We learn from Figs.~\ref{fig:two} and \ref{fig:three} that $\Delta$ is
insignificant, below 0.02\% in size, for all considered values of $\tan\beta$
and $m_A$.
We stress that this happens by virtue of the $G_F$ scheme.

In summary, we derived analytic results for the sfermion-induced radiative 
corrections to the cross sections of
$e^+e^-\to\gamma\gamma,\gamma Z,ZZ,W^+W^-$ at one loop in the MSSM and 
presented a phenomenological discussion for the most interesting case,
$e^+e^-\to W^+W^-$, adopting a SUGRA-inspired scenario.
In the latter case, the correction can essentially be quenched by adopting 
the $G_F$ scheme, which could not be anticipated without explicit calculation.
On the other hand, the sfermions are likely to generate the bulk of the MSSM
correction to $e^+e^-\to W^+W^-$ because of their multiplicities in flavour
and colour.
This expectation is substantiated by a study of the MSSM corrections to the
$\gamma W^+W^-$ and $ZW^+W^-$ TGC's \cite{arh}.
We conclude that significant deviations of the measured cross section of
$e^+e^-\to W^+W^-$ from its SM predictions will not point towards the 
SUGRA-inspired MSSM.

\vspace{1cm}
\noindent
{\bf Note added}
\smallskip

\noindent
After the completion of this work, we received a preprint \cite{ala} which
reports on the MSSM sfermion corrections to the cross section of
$e^+e^-\to W^+W^-$ in the modified minimal-subtraction scheme.
The analytic results for the vector-boson vacuum polarizations and the proper
vertex corrections given in Eqs.~(B.1), (B.3)--(B.5), (B.8) and (B.9) of
Ref.~\cite{ala} agree with our Eqs.~(\ref{eq:pi}) and (\ref{eq:dm}),
respectively.
The agreement was also established numerically to very high precision.

\vspace{1cm}
\noindent
{\bf Acknowledgements}
\smallskip

\noindent
We are grateful to Ralf Hempfling for his collaboration in earlier stages of 
this work.
We thank Jean-Loic Kneur, Gilbert Moultaka, Joannis Papavassiliou, and Jay
Watson for useful comments on Ref.~\cite{arh}.
The work of A.A.B.B. was supported by the Friedrich-Ebert-Stiftung through
Grant No.~219747.
The II. Institut f\"ur Theoretische Physik is supported by the
Bundesministerium f\"ur Bildung und Forschung under Contract No.\ 05~HT9GUA~3,
and by the European Commission through the Research Training Network
{\it Quantum Chromodynamics and the Deep Structure of Elementary Particles}
under Contract No.\ ERBFMRXCT980194.

\newpage

\newpage

\newpage

\end{document}